\documentclass[conference]{IEEEtran}
\usepackage{makeidx}  
\usepackage{latexsym}
\usepackage{amsmath, amssymb, ascmac}
\usepackage{amsfonts, url}
\usepackage{float}

\newtheorem{Definition}{Definition}
\newtheorem{Theorem}{Theorem}      
\newtheorem{Lemma}{Lemma}  
\newtheorem{Remark}{Remark}

\newtheorem{Claim}{Claim}

\makeatletter
\def\@begintheorem#1#2{\trivlist\item[\hskip \labelsep{\bf #1\ #2\normalsize.}]}
\def\@opargbegintheorem#1#2#3{\trivlist\item[\hskip \labelsep{\bf#1\ #2\ (#3)\normalsize.}]}

\def\QED{{\unskip\nobreak\hfil\penalty50
\hskip1em\hbox{}\nobreak\hfil \hbox{$\square$\hskip1em}
\parfillskip\z@ \finalhyphendemerits\z@\par}}
\makeatother

\ifCLASSINFOpdf
\else
\fi
\allowdisplaybreaks

\begin{document}
%
\title{Timed-Release Secret Sharing Scheme with Information Theoretic Security}

\author{\IEEEauthorblockN{Yohei Watanabe and Junji Shikata}
\IEEEauthorblockA{Graduate School of Environment and Information Sciences,
Yokohama National University, Japan\\
Email: watanabe-yohei-xs@ynu.jp, shikata@ynu.ac.jp}
}
%


%


\maketitle

\begin{abstract}
In modern cryptography, the secret sharing scheme is an important cryptographic primitive and it is used in various situations. 
In this paper, a timed-release secret sharing scheme (TR-SS) with information-theoretic security is first studied. TR-SS is a secret sharing scheme with the property that participants more than a threshold number can reconstruct a secret by using their shares only when the time specified by a dealer has come.
Specifically, in this paper we first introduce a model and formalization of security for TR-SS based on the traditional secret sharing scheme and information-theoretic timed-release security. We also derive tight lower bounds on the sizes of shares, time-signals, and entities' secret-keys required for TR-SS. In addition, we propose a direct construction for TR-SS. Our direct construction is optimal in the sense that the construction meets equality in each of our bounds. As a result, it is shown that the timed-release security can be realized without any additional redundancy on the share-size.
\end{abstract}


%
\IEEEpeerreviewmaketitle

\section{Introduction}
Secret sharing schemes were proposed independently by Shamir \cite{Sha79} and Blakley \cite{Bla79}.
In a $(k,n)$-threshold secret sharing ($(k,n)$-SS for short) scheme (e.g. see \cite{Sha79}), a dealer shares a secret among all participants, and then, $k$ participants can reconstruct the secret while any $k-1$ participants obtain no information on the secret. 
Since Shamir and Blakley proposed secret sharing schemes, various research on them have been reported.

On the other hand, ``time'' is intimately related to our lives.
We get up, eat something, do a job, and get asleep at a time of our (or someone's) choice.
From the above reason, it appears that cryptographic protocols associated with ``time'' are useful and meaningful.
Actually, as those protocols, \textit{timed-release cryptographic protocols} introduced in \cite{May93} are well-known. 
 
From the above discussion, it is useful and important to consider a secret sharing scheme with timed-release security.
Therefore, we study such a scheme, which we call a \textit{timed-release secret sharing} (TR-SS) scheme, in this paper.
\medskip

\noindent
\textbf{Timed-Release Security.}
Informally, the goal of timed-release cryptography is {\it to securely send a certain information into the future}. For instance, in timed-release encryption, a sender transmits a ciphertext so that a receiver can decrypt it when the time which the sender specified has come, and the receiver cannot decrypt it before the time.   
The timed-release cryptography was first proposed by May \cite{May93} in 1993, and after that, Rivest et al. \cite{RSW96} developed it in a systematic and formal way.
Since Rivest et al. gave a formal definition of timed-release encryption (TRE) in \cite{RSW96}, various research on timed-release cryptography including timed-release signatures (e.g., \cite{GJ03,GP03}) and timed-release encryption have been done based on computational security. In particular, TRE in the public-key setting has been recently researched on intensively (e.g., \cite{CB05,CLQ05,CHS07}). 
Recently, information-theoretically (or unconditionally) secure timed-release cryptography was proposed by Watanabe et al. \cite{WSS12}.
In addition, they investigated not only an encryption but also a key-agreement and an authentication code with information-theoretic timed-release security.
To the best of our knowledge, however, there is no paper which reports on the study of secret sharing schemes with (information-theoretic) timed-release security.

\medskip

\noindent
\textbf{Our Contribution.}
In adding timed-release functionality to secret sharing schemes, we conceive the following two types of schemes.

One is a secret sharing scheme such that information associated with time (called \textit{time-signals}) is required whenever a secret is reconstructed, which means a secret sharing scheme with a simple combination of traditional secret sharing functionality and timed-release functionality.
For realizing it, we propose $(k,n)$-TR-SS in this paper.
In $(k,n)$-TR-SS, a dealer can specify positive integers $k, n$ with $k \le n$, where $n$ is the number of participants and $k$ is a threshold value, and future time when a secret can be recovered; and the secret can be reconstructed from at least $k$ shares and a time-signal at the specified time.
On the other hand, participants cannot reconstruct the secret without the time-signal even if they can obtain all shares.
Specifically, we define a model and security notions of $(k,n)$-TR-SS, and we derive lower bounds on the sizes of shares, time-signals, and entities' secret-keys required for $(k,n)$-TR-SS.
Moreover, we provide a direct construction of $(k,n)$-TR-SS, which is constructed by using polynomials over finite fields and provably secure in our security definition.
In addition, we show that the direct construction meets the lower bounds on the sizes of shares, time-signals, and entities' secret-keys with equalities. 
Therefore, it turns out that our lower bounds are tight, and that the direct construction is optimal.
%

Another one is a \textit{hybrid} TR-SS, which means a secret sharing scheme in which traditional secret sharing functionality and timed-release functionality are simultaneously realized.
In our hybrid TR-SS, a secret can be reconstructed, if one of the following condition is satisfied: 
a secret can be reconstructed from $k_1$ shares and a time-signal at a specified time as in the $(k_1,n)$-TR-SS; or  
a secret can be reconstructed from $k_2$ shares as in the traditional $(k_2,n)$-SS. Hence, we consider two threshold values $k_1,k_2$ to define a model of the hybrid TR-SS, and we propose $(k_1,k_2,n)$-TR-SS as such a model, where $k_1\le k_2 \le n$.
Specifically, in $(k_1,k_2,n)$-TR-SS, a dealer can specify future time, and arbitrarily chooses $k_1$, $k_2$ and $n$.
At least $k_1$ (and less than $k_2$) participants can reconstruct a secret with a time-signal at the specified time, and at least $k_2$ participants can reconstruct a secret \textit{without} any time-signal (i.e. they can reconstruct from \textit{only} their shares).
Specifically, we define a model and security notions of $(k_1,k_2,n)$-TR-SS, and we derive {\it tight} lower bounds on the sizes of shares, time-signals, and entities' secret-keys required for $(k_1,k_2,n)$-TR-SS.
Moreover, we provide two direct constructions of $(k_1,k_2,n)$-TR-SS:
One is a naive construction, which is very simple, however, does not meet the above lower bounds with equalities;
The other is an {\it optimal} construction, which meets the above lower bounds with equalities.
%

In particular, a theoretically-interesting point in our results includes that the timed-release security can be realized without any additional redundancy on the share-size in both schemes.

\medskip

\noindent
\textbf{Applications of TR-SS.}
Our TR-SS is a secret sharing scheme with timed-release property. 
We consider one of particular applications of TR-SS. 
Recently, in a real world setting, secret sharing schemes have been considered as applications, especially for cloud computing (e.g., secure data storage services).
As represented by big data, information sharing via cloud computing has been developing over recent years.
By applying TR-SS, an information provider can specify arbitrary time when the information is shared.
Actually, the following case is known:
Some companies share their big data, which usually includes sensitive data, and that each company uses shared data for its own business.
Then, by using TR-SS, each company can specify future time when other companies can use such sensitive information.
Therefore, we can say that TR-SS can provide more \textit{flexible} security than traditional secret sharing schemes.

Furthermore, TR-SS can also provide cryptographic protocols with timed-release functionality.
For example, we can construct information-theoretically secure TRE in the two-user setting from $(1,1)$-TR-SS and the one-time pad as follows.
For a plaintext $M$ and a shared key $K$, a sender chooses a random number $r$ whose length is equal to the plaintext-length, and computes a cipertext $C:=M\oplus r \oplus K$.
Then, the sender specifies future time, and he generates one share from the secret $r$ by $(1,1)$-TR-SS.
A receiver can compute $C\oplus K=M\oplus r$ by using the shared key $K$ in advance, however, he cannot obtain $M$ until the specified time comes since he can get $r$ only after the specified time. 
In a similar way, it is expected that TR-SS is useful for building other timed-release cryptographic protocols such as timed-release authentication code \cite{WSS12} in the two-user setting, and that TR-SS might be able to provide some new timed-release cryptographic protocols, e.g., timed-release threshold encryption.

\medskip
\noindent
\textbf{Organization of this paper.}
The rest of this paper is organized as follows. 
In Sections \ref{Sec_TR-SS1} and \ref{Sec_TR-SS2}, we describe $(k,n)$-TR-SS and $(k_1,k_2,n)$-TR-SS, respectively, which are based on the ideas according to \cite{Sha79,KGH83,WSS12}.
Specifically, in each section, we define a model and security of each scheme, and derive lower bounds on the sizes of shares, time-signals and secret-keys required for each scheme, respectively.
Furthermore, we propose a direct construction of each scheme, and show it is provably secure and optimal. 
Finally, in Section \ref{Conclusion}, we give concluding remarks of this paper.

\medskip
\noindent
\textbf{Notation.}
Throughout this paper, we use the following notation. 
Generally speaking, $X$ indicates a random variable which takes values in $\mathcal{X}$ (e.g., $A, B,$ and $C$ are random variables which take values in $\mathcal{A}, \mathcal{B},$ and $\mathcal{C}$, respectively).
For any finite set $\mathcal{Z}$ and arbitrary non-negative integers $z_1,z_2$, let $\mathcal{PS}(\mathcal{Z}, z_1, z_2) := \{ Z \subset \mathcal{Z} | z_1 \le |Z| \le z_2 \}$ be the family of all subsets of $\mathcal{Z}$ whose cardinality is at least $z_1$ but no more than $z_2$.

\section{$(k,n)$-Timed-Release Secret Sharing Scheme}\label{Sec_TR-SS1}
In this section, we propose a model and a security definition of $(k,n)$-TR-SS.
In $(k,n)$-TR-SS, a time-signal at the specified time is always required when a secret is reconstructed.
In other words, a secret cannot be reconstructed without a time-signal at the specified time even if there are all shares.

\subsection{The Model and Security Definition} \label{Sec_Model1}
First, we introduce the model of $(k, n)$-TR-SS.
Unlike traditional secret sharing schemes \cite{Bla79,Sha79}, we assume that there is a trusted authority (or a trusted initializer) $TA$ whose role is to generate and to distribute secret-keys of entities. We call this model the \textit{trusted initializer model} as in \cite{Riv99}.%
In $(k, n)$-TR-SS, there are $n + 3$ entities, a dealer $D$, $n$ participants $P_1, P_2, \ldots, P_n$, a time-server $TS$ for broadcasting time-signals at most $\tau$ times and a trusted initializer $TA$, where $k$, $n$ and $\tau$ are positive integers.
In this paper, we assume that the identity of each user $P_i$ is also denoted by $P_i$.

Informally, $(k, n)$-TR-SS is executed as follows.
First, $TA$ generates secret-keys on behalf of $D$ and $TS$.
After distributing these keys via secure channels, $TA$ deletes it in his memory.
Next, $D$ specifies future time, as $D$ wants, when a secret is reconstructed by participants, and he generates $n$ shares from the secret by using his secret-key. 
And, $D$ sends each share to each participant respectively via secure channels.
The time-server $TS$ periodically broadcasts a time-signal which is generated by using his secret-key.
When the specified time has come, at least $k$ participants can compute the secret by using their shares and the time-signal of the specified time.

Formally, we give the definition of $(k, n)$-TR-SS as follows.
In this model, let $\mathcal{P}:=\{P_1, P_2, \ldots, P_n\}$ be a set of all participants.
And also, $\mathcal{S}$ is a set of possible secrets with a probability distribution $P_S$, and $\mathcal{SK}$ is a set of possible secret-keys. $\mathcal{T} := \{ 1, 2, \ldots, \tau \}$ is a set of time.
Let $\mathcal{U}_{i}^{(t)}$ be the set of possible $P_i$'s shares at the time $t\in {\cal T}$. 
Also, $\mathcal{U}_i:=\bigcup_{t=1}^{\tau}\mathcal{U}_i^{(t)}$ is a set of possible $P_i$'s shares for every $i\in \{1,2,\ldots,n \}$, and let $\mathcal{U} := \bigcup^{n}_{i=1}\mathcal{U}_i$. 
In addition, $\mathcal{TI}^{(t)}$ is a set of time-signals at time $t$, and let $\mathcal{TI}:=\bigcup_{t=1}^{\tau}\mathcal{TI}^{(t)}$.
Furthermore, for any subset of participants $\mathcal{J} = \{P_{i_1},\ldots,P_{i_j}\} \subset \mathcal{P}$,  $\mathcal{U}_{\mathcal{J}}^{(t)} := \mathcal{U}_{i_1}^{(t)} \times \cdots \times \mathcal{U}_{i_j}^{(t)}$ denotes the set of possible shares held by $\mathcal{J}$.

\begin{Definition}[$(k,n)$-TR-SS] \label{TR-SS1}
{\it  
A $(k, n)$-timed-release secret sharing scheme ($(k, n)$-TR-SS) $\Pi$ involves $n+3$ entities, $TA, D, P_1,\ldots, P_n,$ and $TS$, and consists of four phases, \textit{Initialize, Extract, Share,} and \textit{Reconstruct}, and five finite spaces, $\mathcal{S}, \mathcal{SK}, \mathcal{U}, \mathcal{T}$, and $\mathcal{TI}$. $\Pi$ is executed based on the above phases as follows.
}
\begin{enumerate}
\item[a)] \textit{Initialize.}
$TA$ generates a secret-key $sk \in \mathcal{SK}$ for $TS$ and $D$. These keys are distributed to corresponding entities via secure channels.
After distributing these keys, $TA$ deletes them from his memory. And, $D$ and $TS$ keep their keys secret, respectively.\footnote{If we consider a situation in which $TS$ is trusted and $TS$ has functionality of generating keys and distributing them to participants by secure private channels, we can identify $TA$ with $TS$ in the situation. However, there may be a situation in which the roles of $TA$ and $TS$ are quite different (e.g., $TA$ is a provider of secure data storage service and $TS$ is a time-signal broadcasting server). Therefore, we assume two entities $TA$ and $TS$ in our model to capture various situations.}
\item[b)] \textit{Share.}
A dealer $D$ randomly selects a secret $s \in \mathcal{S}$ according to $P_S$, and chooses $k$ and $n$.
If $D$ wants the secret $s$ to be reconstructed by participants at future time $t \in \mathcal{T}$, on input the secret $s \in \mathcal{S}$, specified time $t \in \mathcal{T}$ and a secret-key $sk$, $D$ computes a share $u_{i}^{(t)} \in \mathcal{U}_i^{(t)}$ for every $P_i$ $(i=1, 2, \ldots, n)$. And then, $D$ sends a pair of the share and specified time, $(u_i^{(t)}, t)$, to $P_i$ $(i=1, 2, \ldots, n)$ via a secure channel.%
\footnote{More precisely, there is no need to keep the specified time confidential ($D$ only has to send shares via secure channels).}
\item[c)] \textit{Extract.}
For broadcasting a time-signal at each time $t$, $TS$ generates a time-signal $ts^{(t)} \in \mathcal{TI}^{(t)}$ by using his secret-key $sk$ and time $t \in \mathcal{T}$, where for simplicity we assume that $ts^{(t)}$ is deterministically computed by $t$ and $sk$.
\item[d)] \textit{Reconstruct.} 
At the specified time $t$, any set of participants $\mathcal{A}=\{P_{i_1},\ldots,P_{i_j}\} \in \mathcal{PS}(\mathcal{P},k,n)$ can reconstruct the secret $s$ by using their shares $u_{i_1}^{(t)},\ldots,u_{i_j}^{(t)} \ (k \le j \le n)$ and a time-signal $ts^{(t)}$ at the specified time.
\end{enumerate}
\end{Definition}

In the above model, we assume that $\Pi$ meets the following \textit{correctness} property:
If $D$ correctly completes the phase \textit{Share} and $TS$ correctly completes the phase \textit{Extract}, then, for all possible $i \in \{1,2,\ldots,n\}$, $t \in \mathcal{T}$, $s \in \mathcal{S}$, $u_{i}^{(t)} \in \mathcal{U}_{i}$, and $mk^{(t)} \in \mathcal{TI}^{(t)}$, it hold that any $\mathcal{A} \in \mathcal{PS}(\mathcal{P}, k, n)$ will correctly reconstruct the secret $s$ at the end of phase \textit{Reconstruct}, namely, 
\begin{align*}
H(S \mid U_{\mathcal{A}}^{(t)}, TI^{(t)}) = 0.
\end{align*}

\medskip
Next, we formalize a security definition of $(k, n)$-TR-SS based on the idea of the information-theoretic timed-release security \cite{WSS12} and secret sharing schemes (e.g. see \cite{KGH83}).
In $(k, n)$-TR-SS, we consider the following two kinds of security.
The first security which we consider is basically the same as that of the traditional $(k,n)$-SS: less than $k$ participants cannot obtain any information on a secret. In addition to this, as the second security we want to require that even at least $k$ participants cannot obtain any information on a secret before the specified time comes (i.e., before a time-signal at the specified time is received), since we consider timed-release security in this paper. 
Therefore, we formally define secure $(k, n)$-TR-SS as follows.

\begin{Definition}[Security of $(k,n)$-TR-SS] \label{Security1}
{\it  
Let $\Pi$ be $(k, n)$-TR-SS.
$\Pi$ is said to be \textit{secure} if the following conditions are satisfied:
\begin{enumerate} 
\item[(i)] For any $\mathcal{F} \in \mathcal{PS}(\mathcal{P},1,k-1)$ and any $t \in \mathcal{T}$, it holds that
\begin{align*}
H(S \mid U_{\mathcal{F}}^{(t)}, TI^{(1)}, \dots, TI^{(\tau)}) = H(S). 
\end{align*}
\item[(ii)] For any $\mathcal{A} \in \mathcal{PS}(\mathcal{P}, k, n)$ and any $t \in \mathcal{T}$, it holds that
\begin{align*}
&H(S \mid U_{\mathcal{A}}^{(t)}, TI^{(1)}, \dots, TI^{(t-1)}, TI^{(t+1)}, \dots, TI^{(\tau)}) \\
&=H(S). 
\end{align*}
\end{enumerate}
}
\end{Definition}

Intuitively, the meaning of two conditions (i) and (ii) in Definition \ref{Security1} is explained as follows.  
(i) No information on a secret is obtained by any set of less than $k$ participants, even if they obtain time-signals at all the time; 
(ii) No information on a secret is obtained by any set of more than $k-1$ participants, even if they obtain time-signals at all the time except the specified time.\footnote{In this sense, we have formalized the security notion stronger than the security that any set of more than $k-1$ participants cannot obtain any information on a secret {\it before} the specified time, as is the same approach considered in \cite{WSS12}.}

\begin{Remark}
We can also consider the following security definition (the condition (iii)) instead of (i):
No information on a secret is obtained by collusion of $TS$ and any set of less than $k$ participants, namely, this is defined as follows.
\begin{enumerate} {\it
\item[(iii)] For any $\mathcal{F} \in \mathcal{PS}(\mathcal{P}, 1,k-1)$ and for any $t \in \mathcal{T}$, it holds that 
\begin{align*}
H(S \mid U_{\mathcal{F}}^{(t)}, SK) = H(S). 
\end{align*} }
\end{enumerate}
Note that the condition (iii) is stronger than (i). However, we do not consider (iii) in this paper because of the following two reasons: first, the condition (i) is more natural than (iii), since it does not seem natural to consider the situation that any set of less than $k$ participants colludes with $TS$ in the real world; and secondly, our lower bounds in Theorem \ref{Bounds} are still valid even under the conditions (ii) and (iii), in other words, even if we consider the conditions (ii) and (iii), we can derive the same lower bounds in Theorem \ref{Bounds1} since Definition \ref{Security1} is weaker. Interestingly, our direct construction in Section \ref{Direct} also satisfies (iii), and {\it tightness} of our lower bounds and {\it optimality} of our direct construction will be valid not depending on the choice of the condition (i) or (iii). Furthermore, we do not have to consider an attack by dishonest $TS$ only, since $TS$'s master-key is generated independently of a secret. 
\end{Remark}

\subsection{Lower Bounds}\label{Sec_Bounds1}
In this section, we show lower bounds on sizes of shares, time-signals, and secret-keys required for secure $(k,n)$-TR-SS as follows.

\begin{Theorem} \label{Bounds1} {\it
Let $\Pi$ be any secure $(k,n)$-TR-SS. Then, for any $i \in \{1,2,\dots, n\}$ and for any $t \in \mathcal{T}$, we have
\begin{align*}
(i)& \ H(U_i^{(t)}) \ge H(S), & (ii)& \ H(TI^{(t)}) \ge H(S), \\
(iii)& \ H(SK) \ge \tau H(S). &&
\end{align*}
}
\end{Theorem}
The proof follows from the following lemmas.

\begin{Lemma} \label{LemmaU1} {\it
$H(U_i^{(t)}) \ge H(S)$ for any $i \in \{1,2,\dots, n\}$ and any $t \in \mathcal{T}$. }
\end{Lemma}
\textit{Proof.}
The proof of this lemma can be proved in a way similar to the proof in \cite[Theorem 1]{KGH83}.
For arbitrary $i\in \{1,2,\dots,n\}$, we take a subset $\mathcal{B}_i \in \mathcal{PS}(\mathcal{P}\setminus\{P_i\},k-1,k-1)$ of participants.
Then, for any $t \in \mathcal{T}$, we have
\begin{flalign}
H(U_i^{(t)}) 
&\ge H(U_i^{(t)} \mid U_{\mathcal{B}_i}^{(t)}, TI^{(t)}) \nonumber\\ 
&\ge I(S;U_i^{(t)} \mid U_{\mathcal{B}_i}^{(t)}, TI^{(t)}) \nonumber\\
&= H(S \mid U_{\mathcal{B}_i}^{(t)}, TI^{(t)}) \label{LemmaU1_01}\\
&= H(S), \label{LemmaU1_02}
\end{flalign}
where (\ref{LemmaU1_01}) follows from the correctness of $(k,n)$-TR-SS and (\ref{LemmaU1_02}) follows from the condition (i) in Definition \ref{Security1}. \QED

\begin{Lemma} \label{LemmaTI1}{\it
$H(TI^{(t)} \mid TI^{(1)}, \dots, TI^{(t-1)}) \ge H(S)$ for any $t \in \mathcal{T}$. In particular, $H(TI^{(t)}) \ge H(S)$ for any $t \in \mathcal{T}$. }
\end{Lemma}
\textit{Proof.} For any $\mathcal{A} \in \mathcal{PS}(\mathcal{P},k,n)$ and any $t \in \mathcal{T}$, we have
\begin{align}
H(TI^{(t)})
\ge&H(TI^{(t)} \mid TI^{(1)}, \dots, TI^{(t-1)}) \nonumber \\
\ge& H(TI^{(t)} \mid U_{\mathcal{A}}^{(t)}, TI^{(1)}, \dots, TI^{(t-1)}) \nonumber \\
\ge& I(S;TI^{(t)} \mid U_{\mathcal{A}}^{(t)}, TI^{(1)}, \dots, TI^{(t-1)}) \nonumber\\
=& H(S \mid U_{\mathcal{A}}^{(t)}, TI^{(1)}, \dots, TI^{(t-1)}) \label{LemmaTI1_01}\\
=& H(S), \label{LemmaTI1_02}
\end{align}
where (\ref{LemmaTI1_01}) follows from the correctness of $(k,n)$-TR-SS and (\ref{LemmaTI1_02}) follows from the condition (ii) in Definition \ref{Security1}. \QED

\begin{Lemma}\label{LemmaSK1} {\it
$H(SK) \ge \tau H(S)$. }
\end{Lemma}
\textit{Proof.} We have 
\begin{flalign*}
H(SK) 
&\ge I(TI^{(1)}, \dots, TI^{(\tau)};SK) \nonumber\\ 
&= H(TI^{(1)}, \dots, TI^{(\tau)}) - H(TI^{(1)}, \dots, TI^{(\tau)} \mid SK) \nonumber\\ 
&= H(TI^{(1)}, \dots, TI^{(\tau)}) \nonumber\\ 
&=\sum_{t=1}^{\tau}H(TI^{(t)} \mid TI^{(1)}, \dots, TI^{(t-1)}) \\
&\ge \tau H(S),
\end{flalign*}
where the last inequality follows from Lemma \ref{LemmaTI1}. \QED 

\medskip
\noindent
\textit{Proof of Theorem \ref{Bounds1}:}
From Lemmas \ref{LemmaU1}-\ref{LemmaSK1}, the proof of Theorem \ref{Bounds1} is completed. \QED

As we will see in Section \ref{Direct}, the above lower bounds are tight since our construction will meet all the above lower bounds with equalities. Therefore, we define optimality of constructions of $(k,n)$-TR-SS as follows.

\begin{Definition} \label{optimality1}
{\it 
A construction of secure $(k,n)$-TR-SS is said to be {\it optimal} if it meets equality in every bound of (i)-(iii) in Theorem \ref{Bounds1}. 
}
\end{Definition}

\begin{Remark}
The secret sharing scheme such that the size of each participant's share is equal to that of the secret is often called an \textit{ideal} secret sharing scheme.
The construction of $(k,n)$-TR-SS in Section \ref{Direct} is optimal, hence, in this sense we achieve \textit{ideal} $(k,n)$-TR-SS.
In terms of share-size, an interesting point is that the timed-release property can be realized without any additional redundancy on the share-size.
Therefore in the sense of the bound on share-size, our results are also regarded as the extension of traditional secret sharing schemes.
\end{Remark}

\subsection{Direct Construction} \label{Direct}
We propose a direct construction of $(k, n)$-TR-SS. 
In addition, it is shown that our construction is optimal. The detail of our construction of $(k, n)$-TR-SS $\Pi$ is given as follows. 
\begin{enumerate}
\item[a)] \textit{Initialize.}
Let $q$ be a prime power, where $q > \max(n, \tau)$, and $\mathbb{F}_q$ be the finite field with $q$ elements. We assume that the identity of each participant $P_i$ is encoded as $P_i \in \mathbb{F}_q \backslash \{0\}$. Also, we assume $\mathcal{T} = \{1, 2, \dots, \tau\} \subset \mathbb{F}_q \backslash \{ 0 \}$ by using appropriate encoding. First, $TA$ chooses uniformly at random $\tau$ distinct numbers $r^{(j)} (1 \le j \le \tau)$ from $\mathbb{F}_q$. $TA$ sends a secret-key $sk:=(r^{(1)},\ldots,r^{(\tau)})$ to $TS$ and $D$ via secure channels, respectively.
\item[b)] \textit{Share.}
First, $D$ chooses a secret $s \in \mathbb{F}_q$. Also, $D$ specifies the time $t$ at which participants can reconstruct the secret.
Next, $D$ randomly chooses a polynomial $f(x) := c^{(t)} + \sum^{k-1}_{i=1}$ $a_{i}x^i$ over $\mathbb{F}_q$, where $c^{(t)}$ is computed by $c^{(t)}:=s+r^{(t)}$ and each coefficient $a_i$ is randomly and uniformly chosen from $\mathbb{F}_q$. 
Finally, $D$ computes $u_i^{(t)} := f(P_i) (i = 1,2,\ldots,n)$ and sends $(u_i^{(t)}, t)$ to $P_i (i=1, 2, \ldots, n)$ via a secure channel.
\item[c)] \textit{Extract.}
For $sk$ and time $t \in \mathcal{T}$, $TS$ broadcasts $t$-th key $r^{(t)}$ as a time-signal at time $t$ to all participants via a (authenticated) broadcast channel.
\item[d)] \textit{Reconstruct.}
First, a set of at least $k$ participants $\mathcal{A}=\{P_{i_1},P_{i_2},\ldots,P_{i_k}\} \in \mathcal{PS}(\mathcal{P}, k,k)$ computes $c^{(t)}$ by Lagrange interpolation:
\begin{align*}
c^{(t)}=\sum_{j=1}^{k}(\prod_{l\neq j}\frac{P_{i_j}}{P_{i_j}-P_{i_l}})f(P_{i_j}),
\end{align*}
from their $k$ shares.
After receiving $ts^{(t)}=r^{(t)}$, they can compute and get $s =c^{(t)} - r^{(t)}$.
\end{enumerate}

The security and optimality of the above construction is stated as follows.

\begin{Theorem} \label{Direct_Security} 
{\it  
The resulting $(k, n)$-TR-SS $\Pi$ by the above construction is secure and optimal. 
}
\end{Theorem}
\textit{Proof.}
First, we show the proof of (i) in Definition \ref{Security1}.
Assume that any $k-1$ participants $\mathcal{F}=\{P_{i_1},\ldots,P_{i_{k-1}}\}\in\mathcal{PS}(\mathcal{P},k-1,k-1)$ try to guess $c^{(t)}$ by using their shares.  Note that they know $r^{(t)} = c^{(t)} - s$ and  
\begin{align*}
f(P_{i_j})=(1,P_{i_j},\ldots,P_{i_j}^{k-1})
\left( 
\begin{array}{c}
c^{(t)} \\
a_1 \\
\vdots \\
a_{k-1}\\ 
\end{array}
\right),
\end{align*}
for $j=1,\ldots,k-1$. Thus, they can know the following matrix:
\begin{align}
\left( 
\begin{array}{cccc}
1 & P_{i_1} & \cdots & P_{i_1}^{k-1} \\
1 & P_{i_2} & \cdots & P_{i_2}^{k-1} \\
\vdots & \vdots & \ddots & \vdots \\
1 & P_{i_{k-1}} & \cdots & P_{i_{k-1}}^{k-1}\\ 
\end{array}
\right)
\left( 
\begin{array}{c}
c^{(t)} \\
a_1 \\
\vdots \\
a_{k-1}\\ 
\end{array}
\right). \label{direct01}
\end{align}
However, from (\ref{direct01}), they cannot guess at least one element of $(c^{(t)}, a_1, \ldots,a_{k-1})$ with probability larger than $1/q$. Therefore, $H(S \mid U_{\mathcal{F}}, TI^{(1)},\ldots,TI^{(\tau)}) = H(S)$ for any $\mathcal{F}\in\mathcal{PS}(\mathcal{P},1,k-1)$ and any $t\in\mathcal{T}$.

Next, we show the proof of (ii) in Definition \ref{Security1}.
Without loss of generality, we suppose that $\tau$ is a specified time, and that all participants try to guess  $r^{(\tau)}$ by using $c^{(\tau)}$ and time-signals at all the time except the time $\tau$, since they obtain $c^{(\tau)}=s+r^{(\tau)}$ from their shares.
They get $\tau-1$ time-signals $r^{(1)},\ldots,r^{(\tau-1)}$.
However, since each time-signal is chosen uniformly at random from $\mathbb{F}_q$, they can guess $r^{(\tau)}$ only with probability $1/q$. By the security of one-time pad, we have $H(S \mid U_{1},\ldots,U_{n}, TI^{(1)},\ldots,T^{(\tau-1)}) = H(S)$. 
Hence, for any $\mathcal{A} \in \mathcal{PS}(\mathcal{P}, k,n)$ and for any $t \in \mathcal{T}$, we have $H(S \mid U_{\mathcal{A}}^{(t)}, TI^{(1)},\ldots,TI^{(t-1)},TI^{(t+1)},\ldots,T^{(\tau)}) = H(S)$.

Finally, it is straightforward to see that the construction satisfies all the equalities of lower bounds in Theorem \ref{Bounds1}. Therefore, the above construction is optimal.
\QED

\section{$(k_1,k_2,n)$-Timed-Release Secret Sharing Scheme}\label{Sec_TR-SS2}
We propose $(k_1,k_2,n)$-TR-SS, where $k_1$ and $k_2$ are threshold values with $1 \le k_1\le k_2 \le n$.
$(k_1,k_2,n)$-TR-SS can realize timed-release functionality---a secret can be reconstructed from at least $k_1$ shares and a time-signal at the specified time---and traditional secret sharing functionality---a secret can be also reconstructed from only at least $k_2$ shares---simultaneously.
In the case that $k=k_1=k_2$, $(k,k,n)$-TR-SS can be considered as traditional $(k,n)$-SS (for details, see Remark \ref{Adequacy}).

\subsection{Model and Security Definition}\label{Model}
In this section, we propose a model and a security definition of $(k_1,k_2,n)$-TR-SS.
First, we introduce a model of $(k_1,k_2,n)$-TR-SS.
In $(k_1,k_2,n)$-TR-SS, there are same entities and sets as those of $(k,n)$-TR-SS.
The main difference from $(k,n)$-TR-SS is that a dealer $D$ can specify two kinds of threshold values, $k_1$ and $k_2$ with $k_1 \le k_2 \le n$:
$k_1$ indicates the number of participants who can reconstruct a secret $s$ with the time-signal at the time specified by the dealer; and $k_2$ indicates the number of participants who can reconstruct $s$ without any time-signals. 
We give the definition of $(k_1,k_2,n)$-TR-SS as follows.

\begin{Definition}[$(k_1,k_2,n)$-TR-SS] \label{TR-SS2}
{\it 
A $(k_1,k_2,n)$-timed-release secret sharing scheme ($(k_1,k_2,n)$-TR-SS) $\Theta$ involves $n+3$ entities, $TA, D, P_1,\ldots, P_n,$ and $TS$, and consists of five phases, \textit{Initialize, Extract, Share, Reconstruct with time-signals} and \textit{Reconstruct without time-signals}, and five finite spaces, $\mathcal{S}, \mathcal{SK}, \mathcal{U}, \mathcal{T}$, and $\mathcal{TI}$. $\Theta$ is executed based on the following phases as follows.
}
\begin{enumerate}
\item[a)] \textit{Initialize.}
This phase follows the same procedure as that of $(k,n)$-TR-SS (see Definition \ref{TR-SS1}).
\item[b)] \textit{Share.}
A dealer $D$ randomly selects a secret $s \in \mathcal{S}$ according to $P_S$.
Then, $D$ chooses $k_1$, $k_2$ and $n$, and specifies future time $t\in\mathcal{T}$ when at least $k_1$ participants can reconstruct $s$.
Then, on input the secret $s$, the specified time $t$ and a secret-key $sk\in\mathcal{SK}$, $D$ computes a share $u_{i}^{(t)} \in \mathcal{U}_i^{(t)}$ for every $P_i$ $(i=1, 2, \ldots, n)$. 
And then, $D$ sends a pair of the share and specified time, $(u_i^{(t)}, t)$, to $P_i$ $(i=1, 2, \ldots, n)$ via a secure channel, respectively.
\item[c)] \textit{Extract.}
This phase follows the same procedure as that of $(k,n)$-TR-SS (see Definition \ref{TR-SS1}).
\item[d)] \textit{Reconstruct with time-signals.} 
At the specified time $t$, any set of participants $\mathcal{A}=\{P_{i_1},\ldots,P_{i_j}\}\in \mathcal{PS}(\mathcal{P},k_1, k_2-1)$ can reconstruct the secret $s$ by using their shares $(u_{i_1}^{(t)},\ldots,u_{i_j}^{(t)})$ $(k_1 \le j < k_2)$ and a time-signal of the specified time $ts^{(t)}$.
\item[e)] \textit{Reconstruct without time-signals.} 
At anytime, any set of participants $\hat{\mathcal{A}}=\{P_{i_1},\ldots,P_{i_j}\}\in \mathcal{PS}(\mathcal{P},k_2, n)$ can reconstruct the secret $s$ by using only their shares $(u_{i_1}^{(t)},\ldots,u_{i_j}^{(t)})$ $(k_2 \le j \le n)$.
\end{enumerate}
\end{Definition}
In the above model, we assume that $\Theta$ meets the following \textit{correctness} properties:
\begin{enumerate}
\item
If $D$ correctly completes the phase \textit{Share} and $TS$ correctly completes the phase \textit{Extract}, then, for all possible $i \in \{1,2,\ldots,n\}$, $t \in \mathcal{T}$, $s \in \mathcal{S}$, $u_{i}^{(t)} \in \mathcal{U}_{i}^{(t)}$, and $ts^{(t)} \in \mathcal{TI}^{(t)}$, it holds that any $\mathcal{A} \in \mathcal{PS}(\mathcal{P},k_1,k_2-1)$ will correctly reconstruct the secret $s$ at the end of phase \textit{Reconstruct with time-signals}, namely,
\begin{align*}
H(S \mid U_{\mathcal{A}}^{(t)}, TI^{(t)}) = 0.
\end{align*}
\item
If $D$ correctly completes the phase \textit{Share}, then, for all possible $i \in \{1,2,\ldots,n\}$, $t \in \mathcal{T}$, $s \in \mathcal{S}$, and $u_{i}^{(t)} \in \mathcal{U}_{i}^{(t)}$, it holds that any $\hat{\mathcal{A}} \in \mathcal{PS}(\mathcal{P},k_2,n)$ will correctly reconstruct the secret $s$ at the end of phase \textit{Reconstruct without time-signals}, namely, 
\begin{align*}
H(S \mid U_{\hat{\mathcal{A}}}^{(t)}) = 0.
\end{align*}
\end{enumerate}

\medskip
Next, we formalize a security definition of $(k_1,k_2,n)$-TR-SS in a similar way to that of $(k,n)$-TR-SS as follows.

\begin{Definition}[Security of $(k_1,k_2,n)$-TR-SS] \label{Security2}
{\it 
Let $\Theta$ be $(k_1,k_2,n)$-TR-SS.
$\Theta$ is said to be \textit{secure} if the following conditions are satisfied:
\begin{enumerate} 
\item[(i)] For any $\mathcal{F} \in \mathcal{PS}(\mathcal{P},1, k_1-1)$ and any $t \in \mathcal{T}$, it holds that
\begin{align*}
H(S \mid U_{\mathcal{F}}^{(t)}, TI^{(1)}, \dots, TI^{(\tau)}) = H(S). 
\end{align*}
\item[(ii)] For any $\hat{\mathcal{F}} \in \mathcal{PS}(\mathcal{P},k_1, k_2-1)$ and any $t \in \mathcal{T}$, it holds that
\begin{align*}
&H(S \mid U_{\hat{\mathcal{F}}}^{(t)}, TI^{(1)}, \dots, TI^{(t-1)},TI^{(t+1)}, \dots, TI^{(\tau)})\\
&=H(S). 
\end{align*}
\end{enumerate}
}
\end{Definition}
In Definition \ref{Security2}, intuitively, the meaning of (i) is the same as that of $(k,n)$-TR-SS (Definition \ref{Security1}), and the meaning of the condition (ii) is explained that no information on a secret is obtained by any set of at least $k_1$ but \textit{no more than $k_2$} participants, even if they obtain time-signals at all the time except the specified time.
We also consider a more strong security notion in a similar to $(k,n)$-TR-SS, however, we do not consider such a strong notion for the same reason as in the case of $(k,n)$-TR-SS.

\begin{Remark} \label{Adequacy}
In the case that $k=k_1=k_2$, the model and security definition of secure $(k,k,n)$-TR-SS (Definitions \ref{TR-SS1} and \ref{Security1}) are the same as those of traditional $(k,n)$-SS.
Namely, our model of $(k_1,k_2,n)$-TR-SS also includes the model of traditional secret sharing schemes.
Therefore, the model and security definition of $(k_1,k_2,n)$-TR-SS can be regarded as a natural extension of those of traditional secret sharing schemes.
\end{Remark}

\subsection{Lower Bounds}\label{Sec_Bounds2}
In this section, we show lower bounds on sizes of shares, time-signals, and secret-keys required for secure $(k_1,k_2,n)$-TR-SS as follows.

\begin{Theorem} \label{Bounds2} {\it
Let $\Theta$ be any secure $(k_1,k_2,n)$-TR-SS. Then, for any $i \in \{1,2,\dots, n\}$ and for any $t \in \mathcal{T}$, we have
\begin{align*}
(i) \ H(U_i^{(t)}) \ge H(S). 
\end{align*}
Moreover, if the above lower bound holds with equality (i.e. $H(U_i^{(t)})=H(S)$ for any $i$ and $t$), we have
\begin{align*}
(ii)& \ H(TI^{(t)}) \ge (k_2-k_1)H(S), &\\
(iii)& \ H(SK) \ge \tau (k_2-k_1) H(S).&
\end{align*}
}
\end{Theorem}
The proof follows from the following lemmas.

\begin{Lemma} \label{LemmaU2} {\it
$H(U_i^{(t)}) \ge H(S)$ for any $i \in \{1,2,\dots, n\}$ and any $t \in \mathcal{T}$. }
\end{Lemma}
\textit{Proof.}
The proof of this lemma can be proved in a way similar to the proof in \cite[Theorem 1]{KGH83}.
For arbitrary $i\in \{1,2,\dots,n\}$, we take a subset $\mathcal{B}_i \in \mathcal{PS}(\mathcal{P}\setminus\{P_i\},k_2-1,k_2-1)$ of participants.
Then, for any $t \in \mathcal{T}$, we have
\begin{flalign}
H(U_i^{(t)}) 
&\ge H(U_i^{(t)} \mid U_{\mathcal{B}_i}^{(t)}, TI^{(1)},\ldots,TI^{(t-1)}) \label{LemmaU2_00}\\ 
&\ge I(S;U_i^{(t)} \mid U_{\mathcal{B}_i}^{(t)}, TI^{(1)},\ldots,TI^{(t-1)}) \nonumber\\
&= H(S \mid U_{\mathcal{B}_i}^{(t)}, TI^{(1)},\ldots,TI^{(t-1)}) \label{LemmaU2_01}\\
&= H(S), \label{LemmaU2_02}
\end{flalign}
where (\ref{LemmaU2_01}) follows from the correctness of $(k_1,k_2,n)$-TR-SS and (\ref{LemmaU2_02}) follows from the condition (ii) in Definition \ref{Security2}. \QED

\begin{Lemma} \label{LemmaTI2}{\it
If $H(U_i^{(t)})=H(S)$ for any $i\in\{1,2,\ldots,n\}$ and $t\in\mathcal{T}$, $H(TI^{(t)}) \ge H(TI^{(t)} \mid TI^{(1)}, \dots, TI^{(t-1)}) \ge (k_2-k_1)H(S)$ for any $t \in \mathcal{T}$. }
\end{Lemma}
\textit{Proof.}
The statement is true in the case that $k_1=k_2$, since Shannon entropy is non-negative.
Therefore, in the following, we assume $k_1<k_2$.
For arbitrary $i\in \{1,2,\dots,n\}$, we take a subset $\mathcal{B}_i \in \mathcal{PS}(\mathcal{P}\setminus\{P_i\},k_2-1,k_2-1)$ of participants.
For any $t \in \mathcal{T}$, we have
\begin{align}
&H(TI^{(t)}) \nonumber \\
\ge&H(TI^{(t)} \mid TI^{(1)}, \dots, TI^{(t-1)}) \nonumber \\
\ge&I(TI^{(t)};U_1^{(t)},U_2^{(t)},\ldots,U_n^{(t)}\mid TI^{(1)},\ldots,TI^{(t-1)}) \nonumber \\
=&H(U_1^{(t)},U_2^{(t)},\ldots,U_n^{(t)}\mid TI^{(1)},\ldots,TI^{(t-1)}) \nonumber \\
&\qquad-H(U_1^{(t)},U_2^{(t)},\ldots,U_n^{(t)}\mid TI^{(1)},\ldots,TI^{(t)}) \nonumber \\
=&H(U_1^{(t)},\ldots,U_{k_1}^{(t)}\mid TI^{(1)},\ldots,TI^{(t-1)}) \nonumber \\
&+H(U_{k_1+1}^{(t)},\ldots,U_{k_2}^{(t)}\mid TI^{(1)},\ldots,TI^{(t-1)},U_1^{(t)},\ldots,U_{k_1}^{(t)}) \nonumber \\
&\quad +H(U_{k_2+1}^{(t)},\ldots,U_n^{(t)}\mid TI^{(1)},\ldots,TI^{(t-1)},U_1^{(t)},\ldots,U_{k_2}^{(t)}) \nonumber \\
&-H(U_1^{(t)},\ldots,U_{k_1}^{(t)}\mid TI^{(1)},\ldots,TI^{(t)}) \nonumber \\
&\quad -H(U_{k_1+1}^{(t)},\ldots,U_{k_2}^{(t)}\mid TI^{(1)},\ldots,TI^{(t)},U_1^{(t)},\ldots,U_{k_1}^{(t)}) \nonumber \\
&\qquad -H(U_{k_2+1}^{(t)},\ldots,U_n^{(t)}\mid TI^{(1)},\ldots,TI^{(t)},U_1^{(t)},\ldots,U_{k_2}^{(t)}) \nonumber \\
\ge&H(U_1^{(t)},\ldots,U_{k_1}^{(t)}\mid TI^{(1)},\ldots,TI^{(t)}) \nonumber \\
&+H(U_{k_1+1}^{(t)},\ldots,U_{k_2}^{(t)}\mid TI^{(1)},\ldots,TI^{(t-1)},U_1^{(t)},\ldots,U_{k_1}^{(t)}) \nonumber \\
&\quad +H(U_{k_2+1}^{(t)},\ldots,U_n^{(t)}\mid TI^{(1)},\ldots,TI^{(t)},U_1^{(t)},\ldots,U_{k_2}^{(t)}) \nonumber \\
&-H(U_1^{(t)},\ldots,U_{k_1}^{(t)}\mid TI^{(1)},\ldots,TI^{(t)}) \nonumber \\
&\quad -H(U_{k_1+1}^{(t)},\ldots,U_{k_2}^{(t)}\mid TI^{(1)},\ldots,TI^{(t)},U_1^{(t)},\ldots,U_{k_1}^{(t)}) \nonumber \\
&\qquad -H(U_{k_2+1}^{(t)},\ldots,U_n^{(t)}\mid TI^{(1)},\ldots,TI^{(t)},U_1^{(t)},\ldots,U_{k_2}^{(t)}) \nonumber \\
=&H(U_{k_1+1}^{(t)},\ldots,U_{k_2}^{(t)}\mid TI^{(1)},\ldots,TI^{(t-1)},U_1^{(t)},\ldots,U_{k_1}^{(t)}) \nonumber \\
&\quad -H(U_{k_1+1}^{(t)},\ldots,U_{k_2}^{(t)}\mid TI^{(1)},\ldots,TI^{(t)},U_1^{(t)},\ldots,U_{k_1}^{(t)}) \nonumber \\
\ge&\sum_{i=k_1+1}^{k_2}H(U_{i}^{(t)}\mid TI^{(1)},\ldots,TI^{(t-1)},U_{\mathcal{B}_i}^{(t)}) \nonumber \\
&-\sum_{i=k_1+1}^{k_2}H(U_{i}^{(t)}\mid TI^{(1)},\ldots,TI^{(t)},U_1^{(t)},\ldots,U_{i-1}^{(t)}) \nonumber \\
=&(k_2-k_1)H(S), \label{LemmaTI2_01}
\end{align}
where (\ref{LemmaTI2_01}) follows from (\ref{LemmaU2_00}) in the proof of Lemma \ref{LemmaU2}, the assumption of $H(U_i^{(t)})=H(S)$, and the following claim.

\begin{Claim} \label{ClaimTI} {\it
If $k_1<k_2$ and $H(U_i^{(t)})=H(S)$ for any $i\in\{1,2,\ldots,n\}$ and $t\in\mathcal{T}$, $H(U_i^{(t)}\mid U_{\mathcal{A}_i},TI^{(t)})=0$ for any $i\in\{1,2,\ldots,n\}$, any $\mathcal{A}_i\in \mathcal{PS}(\mathcal{P}\setminus\{P_i\},k_1,k_2-1)$, and any $t\in\mathcal{T}$.
}
\end{Claim}
\textit{Proof.}
First, for arbitrary $i\in \{1,2,\ldots,n\}$, we take subsets $\mathcal{B}_i:=\mathcal{PS}(\mathcal{P}\setminus\{P_i\},k_1-1,k_1-1)$ and $\mathcal{A}_i:=\mathcal{PS}(\mathcal{P}\setminus\{P_i\},k_1,k_2-1)$ of participants such that $\mathcal{B}_i \subset \mathcal{A}_i$.
Then, for any $t\in\mathcal{T}$, we have
\begin{align}
&H(U_i^{(t)}) \nonumber \\
\ge&H(U_i^{(t)}\mid U_{\mathcal{B}_i}^{(t)},TI^{(t)}) \nonumber \\
\ge&H(U_i^{(t)}\mid U_{\mathcal{B}_i}^{(t)},TI^{(t)})-H(U_i^{(t)}\mid U_{\mathcal{B}_i}^{(t)},TI^{(t)},S) \label{ClaimTI_00} \\
=&I(U_i^{(t)};S\mid U_{\mathcal{B}_i}^{(t)},TI^{(t)}) \nonumber \\
=&H(S\mid U_{\mathcal{B}_i}^{(t)},TI^{(t)}) -H(S\mid U_{\mathcal{B}_i}^{(t)},U_i^{(t)},TI^{(t)})  \nonumber \\
=&H(S\mid U_{\mathcal{B}_i}^{(t)},TI^{(t)}) \label{ClaimTI_01} \\
=&H(S), \label{ClaimTI_02}
\end{align}
where (\ref{ClaimTI_01}) follows form the correctness of $(k_1,k_2,n)$-TR-SS and (\ref{ClaimTI_02}) follows from the condition (i) in Definition \ref{Security2}.

From (\ref{ClaimTI_00}) and the assumption of $H(U_i^{(t)})=H(S)$, we have
\begin{align*}
&H(U_i^{(t)}\mid U_{\mathcal{B}_i}^{(t)},TI^{(t)}) \\
=& H(U_i^{(t)}\mid U_{\mathcal{B}_i}^{(t)},TI^{(t)})-H(U_i^{(t)}\mid U_{\mathcal{B}_i}^{(t)},TI^{(t)},S).
\end{align*}
Therefore, we have
\begin{align*}
H(U_i^{(t)}\mid U_{\mathcal{B}_i}^{(t)},TI^{(t)},S)=0.
\end{align*}
Hence, we have
\begin{align*}
H(U_i^{(t)}\mid U_{\mathcal{A}_i}^{(t)},TI^{(t)}) 
=&H(U_i^{(t)}\mid U_{\mathcal{A}_i}^{(t)},TI^{(t)},S)\\
\le& H(U_i^{(t)}\mid U_{\mathcal{B}_i}^{(t)},TI^{(t)},S)\\
=&0.
\end{align*}
Since $H(U_i^{(t)}\mid U_{\mathcal{A}_i}^{(t)},TI^{(t)})\ge 0$, we have
$H(U_i^{(t)}\mid U_{\mathcal{A}_i}^{(t)},TI^{(t)})=0$. \QED

\medskip
\noindent
\textit{Proof of Lemma \ref{LemmaTI2}:}
From the above claim, the proof of Lemma \ref{LemmaTI2} is completed. \QED

\begin{Lemma}\label{LemmaSK2} {\it
If $H(U_i^{(t)})=H(S)$ for any $i\in\{1,2,\ldots,n\}$ and $t\in\mathcal{T}$, $H(SK) \ge \tau (k_2-k_1)H(S)$. }
\end{Lemma}
\textit{Proof.} We can prove in a similar way to the proof of Lemma \ref{LemmaSK1}.
We have 
\begin{flalign*}
H(SK) 
&\ge I(TI^{(1)}, \dots, TI^{(\tau)};SK) \\ 
&= H(TI^{(1)}, \dots, TI^{(\tau)}) - H(TI^{(1)}, \dots, TI^{(\tau)} \mid SK) \\ 
&= H(TI^{(1)}, \dots, TI^{(\tau)}) \\ 
&=\sum_{t=1}^{\tau}H(TI^{(t)} \mid TI^{(1)}, \dots, TI^{(t-1)}) \\
&\ge \tau (k_2-k_1)H(S),
\end{flalign*}
where the last inequality follows from Lemma \ref{LemmaTI2}. \QED 

\medskip
\noindent
\textit{Proof of Theorem \ref{Bounds2}:}
From Lemmas \ref{LemmaU2}-\ref{LemmaSK2}, the proof of Theorem \ref{Bounds2} is completed. \QED

As we will see in Section \ref{Optimal}, the lower bounds in Theorem \ref{Bounds2} are tight since our construction will meet all the above lower bounds with equalities. Therefore, we define optimality of constructions of $(k_1,k_2,n)$-TR-SS as follows.

\begin{Definition} \label{optimality2}
{\it 
A construction of secure $(k_1,k_2,n)$-TR-SS is said to be {\it optimal} if it meets equality in every bound of (i)-(iii) in Theorem \ref{Bounds2}. 
}
\end{Definition}

\subsection{Construction} \label{Construction} 
We propose a direct construction of $(k_1,k_2,n)$-TR-SS. 
In addition, it is shown that our construction is optimal.
Before that, we show a naive construction based on $(k_1,n)$-TR-SS and $(k_2,n)$-SS, which is not optimal.

\subsubsection{Naive Construction}
Our idea of a naive construction is a combination of $(k_1,n)$-TR-SS (Section \ref{Direct}) and Shamir's $(k_2,n)$-SS \cite{Sha79}.
\begin{enumerate}
\item[a)] \textit{Initialize.}
Let $q$ be a prime power, where $q > \max(n, \tau)$, and $\mathbb{F}_q$ be the finite field with $q$ elements. We assume that the identity of each participant $P_i$ is encoded as $P_i \in \mathbb{F}_q \backslash \{0\}$. Also, we assume $\mathcal{T} = \{1, 2, \dots, \tau\} \subset \mathbb{F}_q \backslash \{ 0 \}$ by using appropriate encoding. First, $TA$ chooses uniformly at random $\tau$ distinct numbers $r^{(j)} (1 \le j \le \tau)$ from $\mathbb{F}_q$. $TA$ sends a secret-key $sk:=(r^{(1)},\ldots,r^{(\tau)})$ to $TS$ and $D$ via secure channels, respectively.
\item[b)] \textit{Share.}
First, $D$ chooses a secret $s \in \mathbb{F}_q$. Also, $D$ specifies the time $t$ when at least $k_1$ participants can reconstruct the secret and chooses $t$-th key $r^{(t)}$.
Next, $D$ randomly chooses two polynomials $f_1(x) := s + r^{(t)} + \sum^{{k_1}-1}_{i=1}a_{1i}x^i$ and $f_2(x) := s + \sum^{{k_2}-1}_{i=1}a_{2i}x^i$ over $\mathbb{F}_q$, where each coefficient is randomly and uniformly chosen from $\mathbb{F}_q$.
Then, $D$ computes $u_i^{(t)} := (f_1(P_i),f_2(P_i))$.
Finally, $D$ sends $(u_i^{(t)}, t)$ to $P_i (i=1, 2, \ldots, n)$ via a secure channel.
\item[c)] \textit{Extract.}
For $sk$ and time $t \in \mathcal{T}$, $TS$ broadcasts $t$-th key $r^{(t)}$ as a time-signal at time $t$ to all participants via a (authenticated) broadcast channel.
\item[d)] \textit{Reconstruct with time-signals.}
First, $\mathcal{A}=\{P_{i_1},P_{i_2},\ldots,P_{i_{k_1}}\} \in \mathcal{PS}(\mathcal{P},k_1, k_1)$ computes $s + r^{(t)}$ by Lagrange interpolation:
\begin{align*}
s + r^{(t)}=\sum_{j=1}^{k_1}(\prod_{l\neq j}\frac{P_{i_j}}{P_{i_j}-P_{i_l}})f_1(P_{i_j}),
\end{align*}
from $(f_1(P_{i_1}),\ldots,f_1(P_{i_{k_1}}))$.
After receiving $ts^{(t)}=r^{(t)}$, they can compute and get $s =s+ r^{(t)}- ts^{(t)}$.
\item[e)] \textit{Reconstruct without time-signals.}
any $\hat{\mathcal{A}}=\{P_{i_1},P_{i_2},\ldots,P_{i_{k_2}}\} \in \mathcal{PS}(\mathcal{P},k_2,k_2)$ computes
\begin{align*}
s=\sum_{j=1}^{k_2}(\prod_{l\neq j}\frac{P_{i_j}}{P_{i_j}-P_{i_l}})f_2(P_{i_j}),
\end{align*}
by Lagrange interpolation from $(f_2(P_{i_1}),\ldots,f_2(P_{i_{k_2}}))$.
\end{enumerate}

It is easy to see that the above construction is secure, since this construction is a simple combination of $(k_1,n)$-TR-SS and Shamir's $(k_2,n)$-SS.
Also, the above construction is simple, however not optimal since the resulting share-size is twice as large as that of secrets.

\subsubsection{Optimal (but Restricted\protect\footnote{In this optimal construction, a dealer is only allowed to choose $k_1$ and $k_2$ such that $k_2-k_1\le\ell$, where $\ell$ is determined by $TA$ in the phase \textit{Initialize}. In this sense, this construction is restricted.}) Construction} \label{Optimal}
To achieve an optimal construction, we use the technique as in \cite{JS13}:
In the phase \textit{Share}, the dealer computes public parameters, and the public parameters are broadcasted to participants or else stored on a publicly accessible authenticated bulletin board.
The detail of our construction is given as follows.

\begin{enumerate}
\item[a)] \textit{Initialize.}
Let $q$ be a prime power, where $q > \max(n, \tau)$, and $\mathbb{F}_q$ be the finite field with $q$ elements. We assume that the identity of each participant $P_i$ is encoded as $P_i \in \mathbb{F}_q \backslash \{0\}$. Also, we assume $\mathcal{T} = \{1, 2, \dots, \tau\} \subset \mathbb{F}_q \backslash \{ 0 \}$ by using appropriate encoding. First, $TA$ chooses $\ell$, which is the maximum difference between $k_2$ and $k_1$.
Note that $k_1$ and $k_2$ will be determined by a dealer $D$ in the phase \textit{Share}.
Then, $TA$ chooses $\tau\ell$ numbers $r^{(t)}_{i} \ (1\le i \le \ell)$ and $(1\le t\le \tau)$ from $\mathbb{F}_q$ uniformly at random.
$TA$ sends a secret-key $sk:=\{(r^{(t)}_1,r^{(t)}_2,\ldots,r^{(t)}_{\ell})\}_{1\le t \le \tau}$ to $TS$ and $D$ via secure channels, respectively.
\item[b)] \textit{Share.}
First, $D$ randomly selects a secret $s \in \mathbb{F}_q$, and chooses $k_1$, $k_2$ and $n$ such that $k_2-k_1\le\ell$.
Also, $D$ specifies the time $t$ when at least $k_1$ participants can reconstruct the secret.
Next, $D$ randomly chooses a polynomial $f(x) := s + \sum^{{k_2}-1}_{i=1}a_{i}x^i$ over $\mathbb{F}_q$, where each coefficient $a_i$ is randomly and uniformly chosen from $\mathbb{F}_q$.
Then, $D$ computes a share $u_i^{(t)} := f(P_i)$ and a public parameter $p_i^{(t)}:=a_{k_1-1+i}+r_i^{(t)} \ (i = 1,2,\ldots,k_2-k_1)$.
Finally, $D$ sends $(u_i^{(t)}, t)$ to $P_i (i=1, 2, \ldots, n)$ via a secure channel and discloses $(p_1^{(t)},\ldots,p_{k_2-k_1}^{(t)})$.
\item[c)] \textit{Extract.}
For $sk$ and time $t \in \mathcal{T}$, $TS$ broadcasts a time-signal at time $t$, $ts^{(t)} :=(r_1^{(t)},r_2^{(t)},\ldots,r_{\ell}^{(t)})$ to all participants via a (authenticated) broadcast channel.
\item[d)] \textit{Reconstruct with time-signals.}
Suppose that all participants receive $ts^{(t)}=(r_1^{(t)},r_2^{(t)},\ldots,r_{\ell}^{(t)})$.
Let $\mathcal{A}=\{P_{i_1},P_{i_2},\ldots,P_{i_{k_1}}\} \in \mathcal{PS}(\mathcal{P},k_1,k_1)$ be a set of any $k_1$ participants.
First, each $P_{i_j} \in \mathcal{A}$ computes $a_{k_1-1+i}^{(t)}=p_i^{(t)}-r_i^{(t)} \ (i=1,2,\ldots,k_2-k_1)$ and constructs $g(x):=\sum^{k_2-1}_{k_1}a_{i}x^{i}$.
Then, each $P_{i_j}$ computes $h(P_{i_j}):=f(P_{i_j})-g(P_{i_j}) \ (j=1,\ldots,k_1)$ such that $h(x):=s +\sum^{{k_1}-1}_{i=1}a_{i}x^i$.
Then, they compute
\begin{align*}
s=\sum_{j=1}^{k_1}(\prod_{l\neq j}\frac{P_{i_j}}{P_{i_j}-P_{i_l}})h(P_{i_j}),
\end{align*}
by Lagrange interpolation from $(h(P_{i_1}),\ldots,h(P_{i_{k_1}}))$.
\item[e)] \textit{Reconstruct without time-signals.}
any $\hat{\mathcal{A}}=\{P_{i_1},P_{i_2},\ldots,P_{i_{k_2}}\} \in \mathcal{PS}(\mathcal{P},k_2,k_2)$ computes
\begin{align*}
s=\sum_{j=1}^{k_2}(\prod_{l\neq j}\frac{P_{i_j}}{P_{i_j}-P_{i_l}})f(P_{i_j}),
\end{align*}
by Lagrange interpolation from their $k_2$ shares.
\end{enumerate}

The security and optimality of the above construction is stated as follows.

\begin{Theorem} \label{Optimal_Security} 
{\it 
The resulting $(k_1,k_2,n)$-TR-SS $\Theta$ by the above construction is secure. Moreover, it is optimal if $k_2-k_1=\ell$. 
}
\end{Theorem}
\textit{Proof.}
First, we show the proof of (i) in Definition \ref{Security2}.
Assume that $k_1-1$ participants $\mathcal{F}=\{P_{i_1},\ldots,P_{i_{{k_1}-1}}\}\in\mathcal{PS}(\mathcal{P},k_1-1,k_1-1)$ try to guess $s$ by using their shares, public parameters, and all time-signals.
$\mathcal{F}$ can compute $g(x)$ from public parameters and the time-signal at the specified time, hence they can get $h(P_{i_l})=f(P_{i_l})-g(P_{i_l}) \ (l=1,\ldots,{k_1}-1)$.
Thus, they can know the following matrix:
\begin{align}
\left( 
\begin{array}{cccc}
1 & P_{i_1} & \cdots & P_{i_1}^{k_1-1} \\
1 & P_{i_2} & \cdots & P_{i_2}^{k_1-1} \\
\vdots & \vdots & \ddots & \vdots \\
1 & P_{i_{k_1-1}} & \cdots & P_{i_{k_1-1}}^{k_1-1}\\ 
\end{array}
\right)
\left( 
\begin{array}{c}
s \\
a_1 \\
\vdots \\
a_{k-1}\\ 
\end{array}
\right). \label{optimal01}
\end{align}
However, from (\ref{direct01}), they cannot guess at least one element of $(s, a_1, \ldots,a_{k_1-1})$ with probability larger than $1/q$.
Therefore, for any $\mathcal{F} \in \mathcal{PS}(\mathcal{P}, 1,k_1-1)$ and any $t \in \mathcal{T}$, we have $H(S \mid U_{\mathcal{F}}^{(t)}, TI^{(1)},\ldots,TI^{(\tau)}) = H(S)$.

Next, we show the proof of (ii) in Definition \ref{Security2}.
Without loss of generality, we suppose that $\tau$ is a specified time, that $k_2-k_1=\ell$, and that ${k_2}-1$ participants try to guess $s$ by using their shares, public parameters, and time-signals at all the time except the time $\tau$.
First, they cannot guess at least one coefficient of $f(x)$ with probability larger than $1/q$ since the degree of $f(x)$ is at most $k_2-1$.
Therefore, they attempt to guess one of $a_{k_1},\ldots,a_{k_2-1}$ by using their $k_2-1$ shares, public parameters and $\tau-1$ time-signals, since if they obtain any one of these coefficient, they can get $f^*(P_{i_l}) \ (l=1,\ldots,{k_2}-1)$ such that the degree of $f^*(x)$ is $k_2-2$ and reconstruct $s$ by Lagrange interpolation.
They know $\tau-1$ time-signals, however, these time-signals $\{(r_1^{(j)},\ldots,r_{\ell}^{(j)})\}_{1\le j \le \tau-1}$ are independent of the time-signal $(r_1^{(\tau)},\ldots,r_{\ell}^{(\tau)})$ at $\tau$.
Hence, by the security of one-time pad, they cannot guess each $a_{k_1-1+i} \ (=p_i^{(\tau)}-r_i^{(\tau)}) \ (1\le i \le k_2-k_1)$ with probability larger than $1/q$ since each $r_i^{(\tau)}$ is chosen from $\mathbb{F}_q$ uniformly at random.
Therefore, we have $H(S \mid U_{l_1}^{(\tau)},\ldots,U_{l_{{k_2}-1}}^{(\tau)}, TI^{(1)},\ldots,T^{(\tau-1)}) = H(S)$. 
Hence, for any $\mathcal{A} \in \mathcal{PS}(\mathcal{P}, k_1,k_2-1)$ and any $t \in \mathcal{T}$, we have $H(S \mid U_{\mathcal{A}}^{(t)}, TI^{(1)},\ldots,TI^{(t-1)},TI^{(t+1)},\ldots,T^{(\tau)}) = H(S)$.

Finally, if $k_2-k_1=\ell$, it is straightforward to see that the construction satisfies all the equalities of lower bounds in Theorem \ref{Bounds2}. Therefore, the above construction is optimal if $k_2-k_1=\ell$. \QED
 
\section{Concluding Remarks} \label{Conclusion}
In this paper, we first studied two kinds of secret sharing schemes with timed-release security in the information-theoretic setting, $(k,n)$-TR-SS and $(k_1,k_2,n)$-TR-SS.
Specifically, we defined a model and security for each scheme, and derived tight lower bounds on sizes of shares, time-signals, and secret-keys required for each scheme.
Moreover, we respectively proposed optimal direct constructions of both schemes.
These results showed that information-theoretic timed-release security can be realized in secret sharing schemes without any redundancy on share-sizes.

In a similar way, it is expected that information-theoretic timed-release security can be realized for secret sharing schemes with any \textit{access structure} without any redundancy on share-sizes.
It would be also interesting to extend our results to timed-release verifiable secret sharing schemes, and furthermore, to multiparty computation schemes with timed-release security. 

\bibliography{IEEEabrv,ITW2014}



%

\end{document}